\documentclass{ws-procs961x669}            
\begin{document}
\title{Avoiding singularities with propagating torsion}
\author{Luca Fabbri$^*$}
\address{DIME, Universit\`{a} di Genova, Via all'Opera Pia 15, 16145 Genova, Italy\\
$^*$E-mail: luca.fabbri@unige.it}
\begin{abstract}
We consider the torsional completion of the theory of gravity in which the torsion is a propagating axial-vector field interacting with spinor fields: we show how this changes the energy conditions leading to singularity formation being avoided.
\end{abstract}
\keywords{Propagating torsion; torsion gravity; spinors; energy conditions; singularities.}
\bodymatter
\section{Introduction}
In Einstein gravity, the HP theorem is a powerful result stating that when some energy conditions are satisfied then singularity formation is unavoidable. To avoid these singularities, one must have such energy conditions violated. For this purpose we will employ the torsion tensor of a general connection on differentiable manifolds.

When torsion, that is the non-symmetric part of a general connection, is allowed to take its place beside the metric of the space-time, the ensuing physical theory has one more field equation, coupling torsion to the spin of matter (called Sciama-Kibble equations\cite{S, K}), beside the usual field equation, coupling the curvature to the energy of matter (the usual Einstein equations). In such torsional extension of gravitation, the field equation coupling torsion to spin might change the structure of the field equation coupling curvature to energy, eventually changing the energy conditions.

In this paper, we will consider the most general torsional extension, in which the torsion field, being equivalent to an axial-vector, is required to have the dynamics of a Proca Lagrangian. And because a form of matter having both energy and spin is given by the spinor field, our goal is to have also spinors allowed in the model.
\section{Torsion-Gravity with Spinors and Energy Conditions}
In the present work, we will have the most general connection designated by $\Gamma^{\rho}_{\mu\nu}$ and decomposable according to $\Gamma^{\alpha}_{\beta\rho}\!=\!\Lambda^{\alpha}_{\rho\beta}\!+\! Q^{\alpha}_{\phantom{\alpha}\beta\rho}/2$ where $\Lambda^{\rho}_{\mu\nu}$ is the (unique) symmetric (and metric-compatible) connection written in terms of the partial derivatives of the metric, called Levi-Civita connection, and $Q^{\alpha}_{\phantom{\alpha}\beta\rho}$ is the antisymmetric part of the connection, known as torsion tensor: we will take it to be completely antisymmetric, and thus writable as the Hodge dual $Q_{\alpha\sigma\nu}\varepsilon^{\alpha\sigma\nu\mu}\!=\!W^{\mu}$ of an axial-vector.

We notice that because of this decomposition, a similar decomposition can be done for every object defined in terms of the connection. So if we define the covariant derivative with respect to the most general torsionfull connection
\begin{eqnarray}
D_{\mu}V^{\alpha}\!=\!\partial_{\mu}V^{\alpha}\!+\!\Gamma^{\alpha}_{\sigma\mu}V^{\sigma}
\end{eqnarray}
and the covariant derivative with respect to the torsionless connection
\begin{eqnarray}
\nabla_{\mu}V^{\alpha}\!=\!\partial_{\mu}V^{\alpha}\!+\!\Lambda^{\alpha}_{\sigma\mu}V^{\sigma}
\end{eqnarray}
then we have the decomposition
\begin{eqnarray}
&D_{\mu}V^{\alpha}\!=\!\nabla_{\mu}V^{\alpha}
\!+\!\frac{1}{2}Q^{\alpha}_{\phantom{\alpha}\sigma\mu}V^{\sigma}
\end{eqnarray}
showing that the torsionfull covariant derivative can be written as the torsionless covariant derivative plus torsional contributions. An analogous decomposition exists for the curvature of the space-time. Because the Lagrangian of the model shall be written in terms of covariant derivatives and curvatures, the above decomposition means that any Lagrangian written with torsionfull quantities is equivalent to the Lagrangian written with torsionless quantities as long as explicit torsional contributions are added as extra terms. And because torsion is the dual of an axial-vector, these extra contributions shall be those pertaining to an axial-vector field.

The dynamics of axial-vector fields, or vector fields more in general, is determined by the Proca Lagrangian. With some insight, it is helpful to introduce the curl $(\partial W)_{\alpha\nu}\!:=\!\partial_{[\alpha}W_{\nu]}\!\equiv\!\nabla_{\alpha}W_{\nu}\!-\!\nabla_{\nu}W_{\alpha}$ for a compact writing of dynamical terms.

The gravitational field will be encoded by the Riemann curvature $R^{\sigma}_{\phantom{\sigma}\kappa\alpha\mu}$, from which $R^{\alpha}_{\phantom{\alpha}\mu\alpha\rho}\!=\!R_{\mu\rho}$ and $R^{\alpha}_{\phantom{\alpha}\alpha}\!=\!R$ are known as Ricci curvature tensor and scalar.

As a last element, we introduce the spinor field. We define the Clifford matrices
$\boldsymbol{\gamma}^{i}$ such that $\{\boldsymbol{\gamma}^{i},\boldsymbol{\gamma}^{j}\}\!=\!2\mathbb{I}\eta^{ij}$ and from which $\boldsymbol{\sigma}_{ik}\!=\![\boldsymbol{\gamma}_{i},\boldsymbol{\gamma}_{k}]/4$ are the generators of the complex Lorentz algebra. Identity $2i\boldsymbol{\sigma}_{ab}\!=\!\varepsilon_{abcd}\boldsymbol{\pi}\boldsymbol{\sigma}^{cd}$ implicitly defines $\boldsymbol{\pi}$ as the parity-odd matrix. After exponentiation of the generators one can form the complex Lorentz group and any object $\psi$ transforming according to such a transformation is a spinor field. Upon introduction of the adjoint procedure $\overline{\psi}\!=\!\psi^{\dagger}\boldsymbol{\gamma}^{0}$ one can define
\begin{align}
S^{a}\!=\!\overline{\psi}\boldsymbol{\gamma}^{a}\boldsymbol{\pi}\psi\ \ \ \
\ \ \ \ U^{a}\!=\!\overline{\psi}\boldsymbol{\gamma}^{a}\psi\label{vectors}\\
\Theta\!=\!i\overline{\psi}\boldsymbol{\pi}\psi\ \ \ \
\ \ \ \ \ \ \ \ \Phi\!=\!\overline{\psi}\psi\label{scalars},
\end{align}
all of which being real tensors. They are such that
\begin{align}
U_{a}U^{a}\!=\!-S_{a}S^{a}\!=\!\Theta^{2}\!+\!\Phi^{2}\label{NORM}\\
U_{a}S^{a}\!=\!0\label{ORTHOGONAL},
\end{align}
known as Fierz re-arrangements.

It is always possible to re-parametrize
\begin{align}
\Theta\!=\!2\phi^{2}\sin{\beta}\ \ \ \
\ \ \ \ \ \ \ \ \Phi\!=\!2\phi^{2}\cos{\beta}
\end{align}
in terms of $\beta$ and $\phi$ called chiral angle and density. Calling
\begin{align}
S^{a}\!=\!2\phi^{2}s^{a}\ \ \ \
\ \ \ \ \ \ \ \ U^{a}\!=\!2\phi^{2}u^{a}
\end{align}
the normalized spin and velocity, the identities (\ref{NORM}-\ref{ORTHOGONAL}) reduce to
\begin{align}
u_{a}u^{a}\!=\!-s_{a}s^{a}\!=\!1\\
u_{a}s^{a}\!=\!0
\end{align}
showing that the velocity has $3$ independent components, the $3$ spatial rapidities, whereas the spin has only $2$ independent components, the $2$ angles that, in the rest-frame, its spatial part forms with one given axis. It is possible to demonstrate that one can always define a real tensor $R_{ij\mu}\!=\!-R_{ji\mu}$ such that the covariant derivatives of spin and velocity are proportional to spin and velocity themselves according to
\begin{align}
\nabla_{\mu}s_{\nu}\!=\!s^{\alpha}R_{\alpha\nu\mu}\ \ \ \
\ \ \ \ \nabla_{\mu}u_{\nu}\!=\!u^{\alpha}R_{\alpha\nu\mu}\label{ds-du}:
\end{align}
the tensor $R_{ij\mu}$ is called tensorial connection. As a last element, we introduce
\begin{align}
2Y_{\mu}\!=\!\nabla_{\mu}\beta\!+\!\frac{1}{2}\varepsilon_{\mu\nu\alpha\rho}R^{\nu\alpha\rho}\\
2Z_{\mu}\!=\!\nabla_{\mu}\ln{\phi^{2}}\!+\!R_{\mu\alpha}^{\phantom{\mu\alpha}\alpha}
\end{align}
for our future convenience. The existence of the $R_{ij\mu}$ tensor and the properties (\ref{ds-du}) can be proven in the formalism called polar formulation\cite{Fabbri:2020ypd}.

The Lagrangian of our model is given by
\begin{align}
L\!=\!-\frac{1}{4}(\partial W)^{2}\!+\!\frac{1}{2}M^{2}W^{2}\!-\!R
+\frac{i}{2}(\overline{\psi}\boldsymbol{\gamma}^{\mu}\boldsymbol{\nabla}_{\mu}\psi
\!-\!\boldsymbol{\nabla}_{\mu}\overline{\psi}\boldsymbol{\gamma}^{\mu}\psi)
\!-\!XW_{\nu}S^{\nu}\!-\!m\Phi\label{full}
\end{align}
where $M$ and $m$ are the masses of torsion and the spinor and $X$ is the torsion-spinor coupling constant: the above is the most general Lagrangian that is compatible with the requirement that torsion be a propagating axial-vector Proca field\cite{Fabbri:2014dxa, Fabbri:2018qzy, Fabbri:2018hhv}.

Its variation gives
\begin{align}
\nabla_{\rho}(\partial W)^{\rho\mu}\!+\!M^{2}W^{\mu}\!=\!XS^{\mu}
\end{align}
\begin{align}
\nonumber
R^{\rho\sigma}\!-\!\frac{1}{2}Rg^{\rho\sigma}
\!=\!\frac{1}{2}[\frac{1}{4}(\partial W)^{2}g^{\rho\sigma}
\!-\!(\partial W)^{\sigma\alpha}(\partial W)^{\rho}_{\phantom{\rho}\alpha}
+M^{2}(W^{\rho}W^{\sigma}\!-\!\frac{1}{2}W^{2}g^{\rho\sigma})+\\
+\frac{i}{4}(\overline{\psi}\boldsymbol{\gamma}^{\rho}\boldsymbol{\nabla}^{\sigma}\psi
\!-\!\boldsymbol{\nabla}^{\sigma}\overline{\psi}\boldsymbol{\gamma}^{\rho}\psi
\!+\!\overline{\psi}\boldsymbol{\gamma}^{\sigma}\boldsymbol{\nabla}^{\rho}\psi
\!-\!\boldsymbol{\nabla}^{\rho}\overline{\psi}\boldsymbol{\gamma}^{\sigma}\psi)
-\frac{1}{2}X(W^{\sigma}S^{\rho}\!+\!W^{\rho}S^{\sigma})]
\end{align}
\begin{align}
i\boldsymbol{\gamma}^{\mu}\boldsymbol{\nabla}_{\mu}\psi
\!-\!XW_{\sigma}\boldsymbol{\gamma}^{\sigma}\boldsymbol{\pi}\psi\!-\!m\psi\!=\!0
\end{align}
as torsional, gravitational and material field equations.

In polar variables, the gravitational field equations become
\begin{align}
\nonumber
R^{\rho\sigma}\!-\!\frac{1}{2}Rg^{\rho\sigma}\!=\!\frac{1}{2}[\frac{1}{4}(\partial W)^{2}g^{\rho\sigma}
\!-\!(\partial W)^{\sigma\alpha}(\partial W)^{\rho}_{\phantom{\rho}\alpha}
+M^{2}(W^{\rho}W^{\sigma}\!-\!\frac{1}{2}W^{2}g^{\rho\sigma})+\\
\nonumber
+\phi^{2}[2m\cos{\beta}u^{\rho}u^{\sigma}\!+\!2XW_{\mu}s^{\mu}u^{\rho}u^{\sigma}
\!-\!XW_{\mu}u^{\mu}(s^{\rho}u^{\sigma}\!+\!s^{\sigma}u^{\rho})
\!-\!XW^{\rho}s^{\sigma}\!-\!XW^{\sigma}s^{\rho}-\\
\nonumber
-2Y_{\mu}s^{\mu}u^{\rho}u^{\sigma}
\!+\!Y_{\mu}u^{\mu}(s^{\rho}u^{\sigma}\!+\!s^{\sigma}u^{\rho})
\!+\!Y^{\rho}s^{\sigma}\!+\!Y^{\sigma}s^{\rho}
+Z_{\mu}u_{\pi}s_{\tau}(\varepsilon^{\mu\pi\tau\sigma}u^{\rho}
\!+\!\varepsilon^{\mu\pi\tau\rho}u^{\sigma})-\\
-\frac{1}{4}(R_{\alpha\nu\pi}\varepsilon^{\rho\alpha\nu\pi}g^{\sigma\kappa}
\!+\!R_{\alpha\nu\pi}\varepsilon^{\sigma\alpha\nu\pi}g^{\rho\kappa}
+R_{\alpha\nu}^{\phantom{\alpha\nu}\sigma}\varepsilon^{\rho\alpha\nu\kappa}
\!+\!R_{\alpha\nu}^{\phantom{\alpha\nu}\rho}\varepsilon^{\sigma\alpha\nu\kappa})s_{\kappa}]].
\end{align}

In effective approximation, all mass terms become dominant compared to the dynamical terms. When this occurs, the torsion field equations reduce to
\begin{align}
&M^{2}W^{\mu}\!=\!XS^{\mu}\label{Wgen}:
\end{align}
because of this, it is now possible to have torsion substituted in terms of the spin axial-vector in the gravitational field equations.

When this is done, the gravitational field equations become
\begin{align}
\nonumber
R^{\rho\sigma}\!-\!\frac{1}{2}Rg^{\rho\sigma}\!=\!\frac{1}{2}\phi^{2}[2(m\cos{\beta}\!-\!\phi^{2}X^{2}/M^{2})u^{\rho}u^{\sigma}
\!+\!2\phi^{2}X^{2}/M^{2}(g^{\rho\sigma}\!-\!u^{\rho}u^{\sigma})-\\
\nonumber
-2Y_{\mu}s^{\mu}u^{\rho}u^{\sigma}
\!+\!Y_{\mu}u^{\mu}(s^{\rho}u^{\sigma}\!+\!s^{\sigma}u^{\rho})
\!+\!Y^{\rho}s^{\sigma}\!+\!Y^{\sigma}s^{\rho}
+Z_{\mu}u_{\pi}s_{\tau}(\varepsilon^{\mu\pi\tau\sigma}u^{\rho}
\!+\!\varepsilon^{\mu\pi\tau\rho}u^{\sigma})-\\
-\frac{1}{4}(R_{\alpha\nu\pi}\varepsilon^{\rho\alpha\nu\pi}g^{\sigma\kappa}
\!+\!R_{\alpha\nu\pi}\varepsilon^{\sigma\alpha\nu\pi}g^{\rho\kappa}
+R_{\alpha\nu}^{\phantom{\alpha\nu}\sigma}\varepsilon^{\rho\alpha\nu\kappa}
\!+\!R_{\alpha\nu}^{\phantom{\alpha\nu}\rho}\varepsilon^{\sigma\alpha\nu\kappa})s_{\kappa}]
\label{gravity}.
\end{align}

Once the dynamics is assigned, the problem of singularity formation is addressed by computing the energy conditions. The strongest is given by $R^{\rho\sigma}u_{\rho}u_{\sigma}\!\geqslant\!0$ and so
\begin{align}
m\cos{\beta}\!-\!4\phi^{2}X^{2}/M^{2}-2Y_{\nu}s^{\nu}\!+\!\frac{1}{2}R_{\alpha\nu\sigma}u^{\sigma}s_{\kappa}u_{\rho}\varepsilon^{\alpha\nu\kappa\rho}\!\geqslant\!0.
\end{align}

In the last expression, we have no control over the terms in $Y_{\nu}$ and $R_{\alpha\nu\sigma}$ but we do not need to. Because they are both linear in the spin axial-vector, they cancel for every statistical average of randomly distributed spins, such as the case of the Big Bang or Black Holes (in addition, the Gordon decomposition $\nabla_{\nu}S^{\nu}\!=\!2m\Theta$ allows us to see that for average spin equal to zero also $\Theta$ must vanish and therefore $\beta\!\rightarrow\!0$): this leaves us with the simple
\begin{align}
m\!-\!4\phi^{2}X^{2}/M^{2}\!\geqslant\!0.
\end{align}
This condition, for larger and larger densities, is violated\cite{Fabbri:2023cot}.

In propagating torsion gravity, in the effective approximation, singularities are avoided because the contribution that dominates the energy condition is a pressure with negative sign. Its sign is negative because the positivity of the energy fixes the relative sign between $R$ and $(\partial W)^{2}$ while the reality of the mass fixes the relative sign between $(\partial W)^{2}$ and $M^{2}W^{2}$ and, as a consequence, all signs are locked.
\section{Conclusion}
In this work, we considered Einstein gravity completed with an axial-vector torsion whose propagation was restricted to follow the dynamics of a Proca field.

The most general dynamics had spinors re-formulated in polar form.

We have seen that when torsion is in effective approximation, and spinors distributed to allow spin averaging off, the strong energy condition is no longer valid.

The reason for this fact is that the torsion tensor generates a negative pressure in the gravitational field equations: such a pressure is always negative because, as torsion is a propagating field, it must have real and positive mass and energy.

Because of this, all relative signs in the Lagrangian were fixed, and those signs remain fixed also in effective approximation, where the torsionally-induced spin-spin interaction results into a pressure. As a consequence, such a pressure is always negative, the energy conditions are violated, and singularities no longer occur.

\section{Acknowledgments}
This work was funded by Next Generation EU project ``Geometrical and Topological effects on Quantum Matter (GeTOnQuaM)''.


\end{document}